\documentclass[12pt, letterpaper, onecolumn]{article}
\pdfoutput=1 

\usepackage[utf8]{inputenc}
\usepackage[dvipsnames]{xcolor}

\usepackage[top=1in, bottom=1in, left=1in, right=1in]{geometry}

\usepackage{graphicx}
\usepackage[scaled]{helvet}
\usepackage{boldmath}
\usepackage{amsmath, amsthm, amssymb}
\usepackage{bm}
\usepackage{array, booktabs, multirow, tabularx}
\usepackage{setspace}
\usepackage{xspace}
\usepackage{ulem}
\usepackage{hyperref}

\usepackage{color}
\usepackage{xcolor}
\usepackage[toc,acronym,section=section,nonumberlist,style=super3col]{glossaries}
\usepackage[nooneline,labelsep=period, font=footnotesize]{caption}
\usepackage[]{float}

\glsdisablehyper
\newacronym{snr}{SNR}{signal-to-noise ratio}

\newcommand{\bs}{\boldsymbol}

\begin{document}

	\date{}
	
	\title{\Large{\bf Deep learning-guided weighted averaging for signal dropout compensation in diffusion-weighted imaging of the liver}}
	\author{Fasil Gadjimuradov$^{1,2}$, Thomas Benkert$^{2}$, Marcel Dominik Nickel$^{2}$, \\Tobit Führes$^3$, Marc Saake$^3$, Andreas Maier$^{1}$}
	
	\maketitle
	\thispagestyle{empty}
	
	\begin{itemize}
	\item[1] Pattern Recognition Lab, Department of Computer Science, Friedrich-Alexander-\\Universität Erlangen-Nürnberg, Erlangen, Germany
	\item[2] MR Application Predevelopment, Siemens Healthcare GmbH, Erlangen, Germany
	\item[3] Institute of Radiology, University Hospital Erlangen, Friedrich-Alexander-Universität Erlangen-Nürnberg, Erlangen, Germany
	\end{itemize}
	
	\vfill	
	
	\noindent \begin{tabular*}{\textwidth}{ll}
		Correspondence to: &Fasil Gadjimuradov\\
		&Pattern Recognition Lab\\ 
		&Department of Computer Science\\
		&Friedrich-Alexander-Universität Erlangen-Nürnberg\\
		&Martensstr. 3, D-91058 Erlangen, Germany\\
		&E-mail: fasil.gadjimuradov@fau.de
	\end{tabular*} \vspace{.5cm}
	
	\noindent Preliminary data of this work was submitted to the Joint Annual Meeting ISMRM-ESMRMB 2022, London, United Kingdom. \vspace{.5cm}

	\noindent
	Number of words (abstract): 247\\
	Number of words (body): approx. 4,910\\
	Number of figures: 10\\
	Number of tables: 0\\
	Number of references: 17\vspace{1cm}
	
	\noindent \textbf{Submitted to Magnetic Resonance in Medicine}

	\linespread{1.5}	
	
	\clearpage	
	\section*{Abstract}
	\small
	\vspace{0.3cm}
	\noindent
	\textit{\textbf{Purpose:}} To develop an algorithm for the retrospective correction of signal dropout artifacts in abdominal diffusion-weighted imaging (DWI) resulting from cardiac motion.
		
	\noindent
	\textit{\textbf{Methods:}} Given a set of image repetitions for a slice, a locally adaptive weighted averaging is proposed which aims to suppress the contribution of image regions affected by signal dropouts. Corresponding weight maps were estimated by a sliding-window algorithm which analyzed signal deviations from a patch-wise reference. In order to ensure the computation of a robust reference, repetitions were filtered by a classifier that was trained to detect images corrupted by signal dropouts. The proposed method, named \textit{Deep Learning-guided Adaptive Weighted Averaging} (DLAWA), was evaluated in terms of dropout suppression capability, bias reduction in the Apparent Diffusion Coefficient (ADC) and noise characteristics.
	
	\noindent
	\textit{\textbf{Results:}} In the case of uniform averaging, motion-related dropouts caused signal attenuation and ADC overestimation in parts of the liver with the left lobe being affected particularly. Both effects could be substantially mitigated by DLAWA while preventing global penalties with respect to signal-to-noise ratio (SNR) due to local signal suppression. Performing evaluations on patient data, the capability to recover lesions concealed by signal dropouts was demonstrated as well. Further, DLAWA allowed for transparent control of the trade-off between SNR and signal dropout suppression by means of a few hyperparameters.
	
	\noindent
	\textit{\textbf{Conclusion:}} This work presents an effective and flexible method for the local compensation of signal dropouts resulting from motion and pulsation. Since DLAWA follows a retrospective approach, no changes to the acquisition are required. \vspace{1cm}
		
	\noindent 	
	\textit{\textbf{Key words}}: diffusion-weighted imaging, liver MRI, cardiac motion artifact, deep learning, liver oncology
	
	\clearpage
	\section{Introduction} \label{sec:introduction}

\begin{figure}[tb]
	\centering
	\includegraphics[width=.85\textwidth]{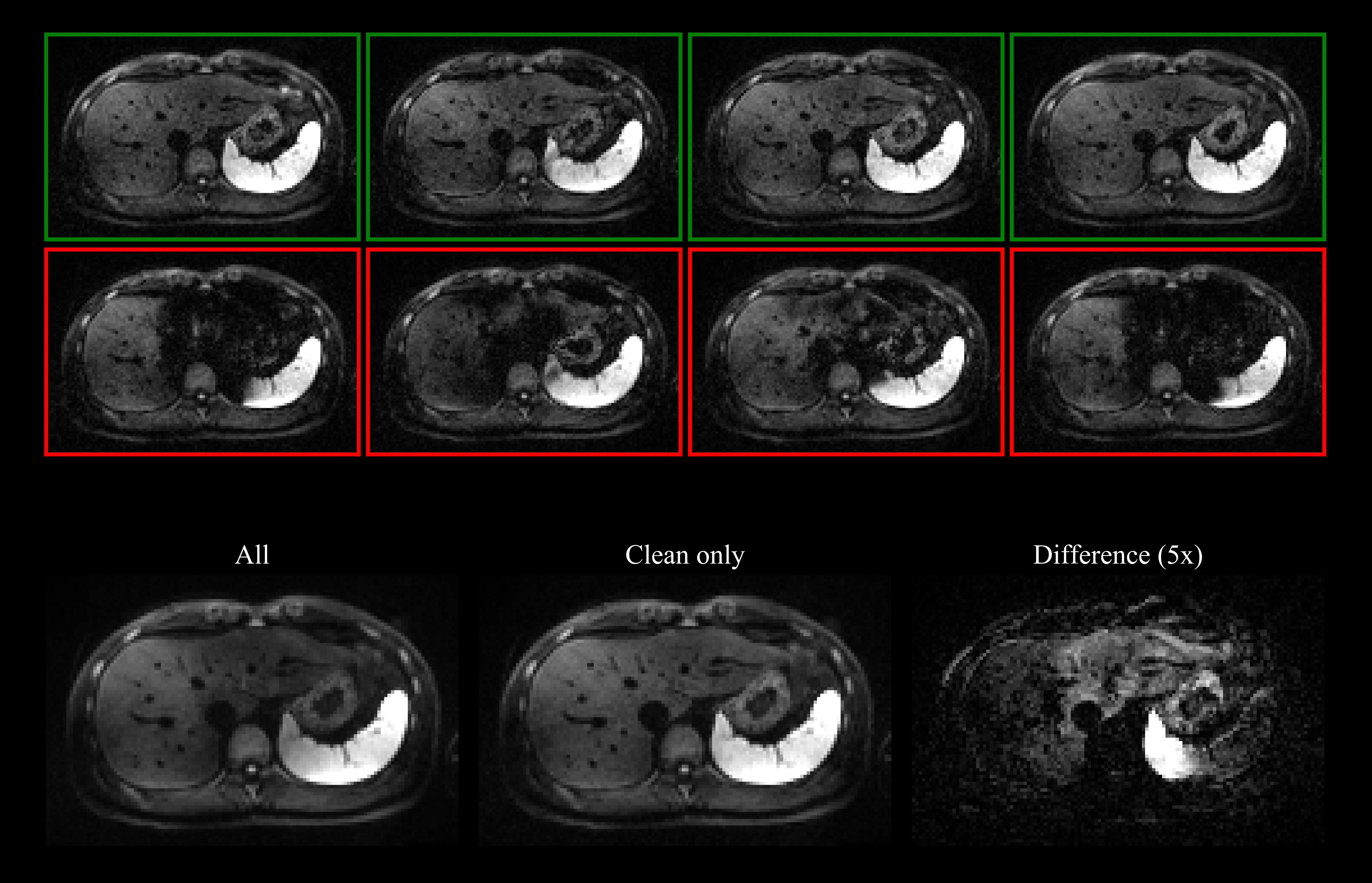}
	\caption{Example of signal dropouts in liver DWI caused by cardiac motion. The first two rows show eight repetitions of an axial liver slice acquired at a $b$-value of 800 s/mm$^2$. The images of the second row are affected by signal dropouts (red frames) leading to the disappearance of large parts of the spleen and the liver, especially the left lobe. When averaging all eight repetitions, the resulting image (``All") suffers from signal attenuation in the left lobe and the neighboring part of the spleen. In contrast, by averaging only the four repetitions from the top row which are marked as uncorrupted (green frames), signal becomes more homogeneous within the liver and the spleen, respectively. However, using a reduced number of repetitions for averaging ``Clean only") comes at the cost of decreased SNR.}
	\label{fig:dropout_example}
	\end{figure}

Diffusion-weighted imaging (DWI) is an integral part of many clinical protocols as it allows to visualize regions of restricted diffusion. In the context of liver imaging, DWI has been shown to be valuable with respect to lesion detection and characterization, assessment of tumor treatment response as well as diagnosis of fibrosis and cirrhosis \cite{LiverDWI1, LiverDWI2, LiverDWI3, Hernando}. Despite its diagnostic value, liver DWI suffers from several technical limitations. As a result of signal decay due to strong diffusion encoding gradients, signal-to-noise ratio (SNR) is inherently low. In practice, this is alleviated by acquiring and averaging multiple image repetitions for every slice. Further, physiological motion constitutes another major challenge in liver DWI. While the duration of single-shot EPI acquisitions is typically short enough to be insensitive to respiratory and bulk motion, rapid cardiovascular motion and pulsation during diffusion encoding can cause intra-voxel phase dispersion which becomes apparent as signal dropout. Due to its proximity to the heart, the left liver lobe is particularly prone to this type of artifact. For typical diffusion encoding schemes the sensitivity to intra-voxel dephasing increases with diffusion weighting, such that a significant portion of repetitions can be affected at high $b$-values. Simple uniform averaging of repetitions may lead to a locally attenuated and inhomogeneous liver signal (see Figure \ref{fig:dropout_example}). As this problem affects low $b$-values significantly less, there is a risk to introduce bias into derived quantitative parameters, such as the Apparent Diffusion Coefficient (ADC).
	
	Several techniques have been proposed to tackle the problem of signal dropouts in abdominal DWI from an acquisition perspective. As motion of the heart is the main source of the artifact, cardiac triggering, as proposed in \cite{Triggering}, constitutes a potential solution. Naturally, it comes with the cost of substantially reducing scan time efficiency and making total scan time difficult to predict. Alternatively, sensitivity to cardiac motion can be reduced via moment nulling of diffusion gradients \cite{Ozaki}. Although more recent methods propose to optimize waveforms with respect to TE (subject to additional sequence, hardware, and $b$-value constraints) \cite{CODE, ODGD, MODI}, increased echo times compared to monopolar waveforms are difficult to prevent. Considering that liver parenchyma has relatively short T2, these methods would nevertheless require a trade-off with respect to SNR. In addition, motion-compensated diffusion encoding is known to lead to unsuppressed signal from blood flow which becomes apparent as bright vessels in DW images and ADC maps. Due to potential resemblance with focal lesions, interpretation of images and quantitative parameters can be therefore confounded. However, previous studies have shown that motion-robust, black-blood liver DWI is possible when using a non-zero but small M1-moment \cite{MODI} or combining velocity-compensated and monopolar waveforms along different diffusion axes \cite{Karampinos}.
	
As an alternative to the prospective methods described above, post-processing techniques for compensating signal dropouts in liver DWI have been investigated as well. In \cite{Liau}, power means were used to adjust the pixel-wise average calculation. By choosing powers greater than 1, pixels with higher signal contribute more to the average, effectively suppressing signal dropouts which are lower in signal by nature. However, the authors noted that a general choice of a power value is nontrivial as it may be specific to subject and $b$-value. A similar weighted averaging approach was employed in \cite{Ichikawa}, where the weight of a pixel was determined as the relative quadratic signal strength. Both works have shown that the problem of signal dropouts can be alleviated to some extent. However, determining weights based on single pixel intensities is highly sensitive to outliers which may lead to excessive over- or under-contributions of certain pixels. Instead of relying on hand-crafted rules for weight estimation, the method presented in \cite{Tamada} employed a neural network to detect corrupted image repetitions which were then entirely excluded from the averaging process. Although this approach was shown to be useful in avoiding effects of signal dropouts in the liver, it is inefficient with respect to SNR as a substantial amount of image information may be discarded. This is especially true when considering that signal dropouts typically affect images locally, leaving the majority of pixels uncorrupted.

The goal of this work is to enable the local suppression of signal dropouts and to minimize SNR penalties in uncorrupted image regions by performing a weighted averaging of image repetitions. In our previous work \cite{Gadji}, weight maps were estimated directly from a set of input repetitions by means of neural network which, however, lacked interpretability. In order to increase the transparency of the method (e.g., it should be plausible why certain pixels are weighted higher or lower) and to let the user have more control over the trade-off between SNR and dropout suppression, a conventional algorithm is proposed which assigns weights based on the amount of deviation from a patch-wise reference value. The algorithm is then complemented by a trained classifier which supports the computation of that reference by marking corrupted repetitions. Realizing a robust solution for retrospective compensation of signal dropouts induced by cardiac motion could foster the accuracy and reproducibility of liver DWI and related quantitative parameters.
	
	\section{Methods} \label{sec:methods}
	
	\subsection{Adaptive Weighting Algorithm}
	\label{sec:AWA}
\begin{figure}[tb]
	\centering
	\includegraphics[width=\textwidth]{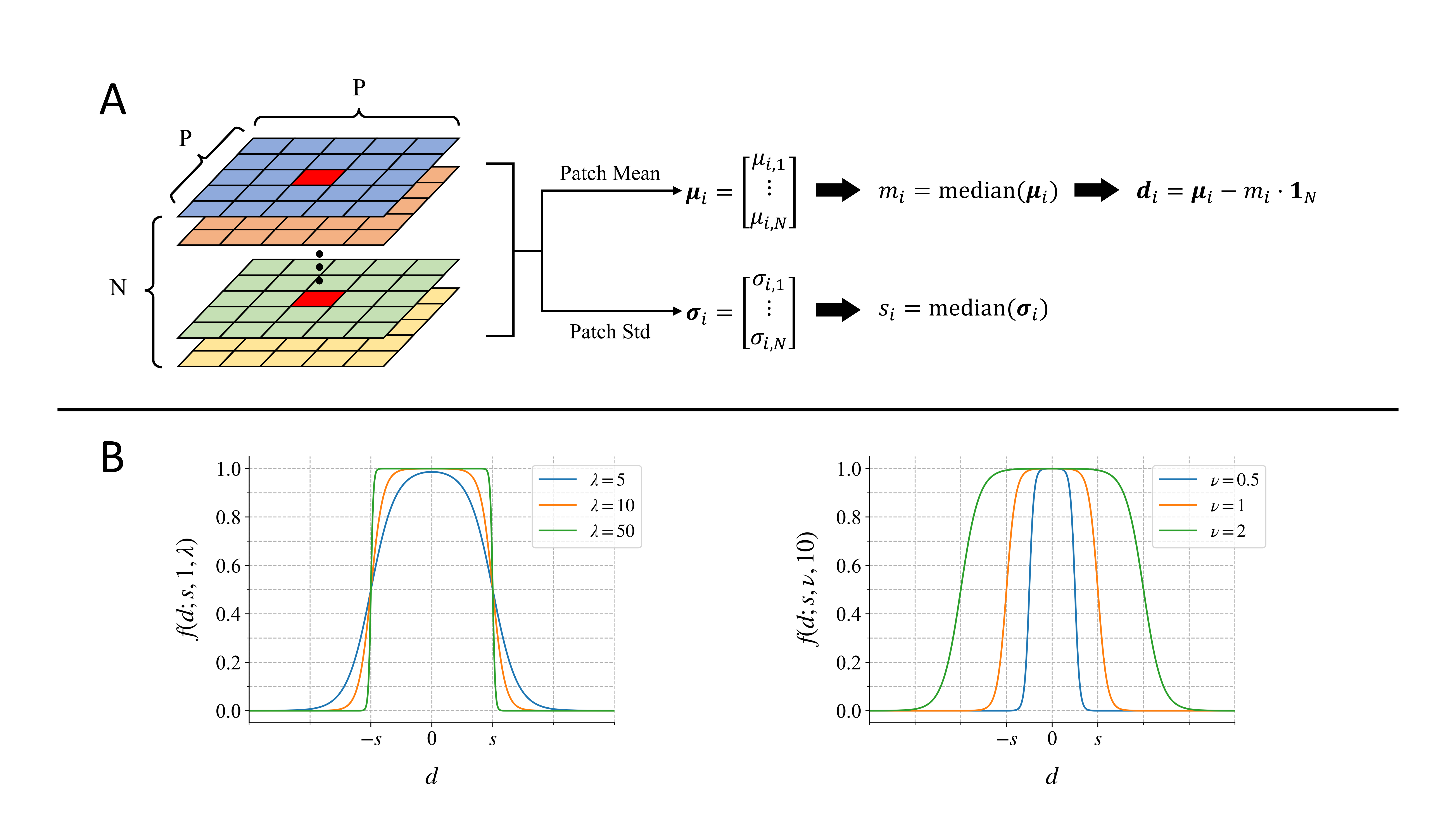}
	\caption{Illustration of the Adaptive Weighting Algorithm (AWA). \textbf{(A)} Patches of size $P \times P$ are extracted at the $i$-th pixel (in red) from all $N$ repetitions with mean $\bs{\mu}_i$ and standard deviation $\bs{\sigma}_i$ of every patch being estimated. Subsequently, both the reference $m_i$ and tolerance value $s_i$ are computed as the median of $\bs{\mu}_i$ and $\bs{\sigma}_i$, respectively. Differences $\bs{d}_i$ with respect to the reference are calculated, accordingly. In addition to hyperparameters $\nu$ and $\lambda$, $\bs{d}_i$ and $s_i$ constitute the input to the weighting function. \textbf{(B)} The left figure shows how $\lambda$ controls the steepness of the function at cut-off points defined by $s_i$. The figure on the right-hand side visualizes the influence of $\nu$ on the width of the tolerance margin.}
	\label{fig:algorithm}
	\end{figure}

The proposed algorithm, referred to as \textit{Adaptive Weighted Averaging} (AWA) in the following, is illustrated in Figure \ref{fig:algorithm}. It is performed in a sliding-window manner by extracting patches of size $P \times P$ from a set of $N$ repetitions at every pixel location $i$. Corresponding patches are represented by their respective estimations of mean $\bs{\mu}_i = [\mu_{i,1}, ..., \mu_{i,N}]^T$ and standard deviation $\bs{\sigma}_i = [\sigma_{i,1},...,\sigma_{i,N}]^T$. The algorithm’s rationale is to compute differences of aggregated patches with respect to some reference value $m_i$ and to penalize corresponding pixels with lower weight if those differences are unlikely to be the result of random image noise. Here, we chose $m_i$ to be the median of patch means $\bs{\mu}_i$ as it is relatively robust against outliers that may represent potential signal dropouts. Differences are calculated as $\bs{d}_i = \bs{\mu}_i - m_i \cdot \bs{1}_N$, accordingly, where $\bs{1}_N$ denotes a vector of ones of length $N$. Based on the assumption that the signal within patches is relatively constant, $\bs{\sigma}_i$ reflects the amount of random noise. A value $s_i$ up to which differences are tolerated is computed as the median of $\bs{\sigma}_i$  which gives a reasonable estimate of the typical signal variation within a patch. Using differences $\bs{d}_i$ and tolerance $s_i$ as arguments, function $f(\cdot)$ computes unnormalized weights $\bs{w}_i = [w_{i,1},...,w_{i,N}]^T$ for the $i$-th pixel of every repetition according to:
\begin{gather}
	w_{i,n} = f(d_{i,n};s_i,\nu,\lambda) = g(d_{i,n};s_i,\nu,\lambda) - g(d_{i,n};-s_i,\nu,\lambda) \quad \text{for} \;\; n=1,...,N \\ 
\text{where} \;\; g(d;s,\nu,\lambda) = \frac{1}{1+e^{-\frac{\lambda}{|\nu s|}(d+\nu s)}} \;. \notag
\end{gather}
By normalizing the estimated weights appropriately, the pixel-wise weighted sum $\widetilde{\rho}_i$ of repetitions $[\rho_{i,1},...,\rho_{i,N}]^T$ can be computed as:
\begin{equation}
	\widetilde{\rho}_i = \sum_{n=1}^N \widetilde{w}_{i,n} \cdot \rho_{i,n} \quad \text{with} \;\; \widetilde{w}_{i,n} = \frac{w_{i,n}}{||\bs{w}_i||} \;.
	\label{eq:weights}
\end{equation}
In particular, $f(\cdot)$ is defined as the smooth approximation of a rectangular function composed of two shifted sigmoid functions $g(\cdot)$. Further, user-defined parameters $\nu$ and $\lambda$ are introduced which allow to control the form of $f(\cdot)$ as shown in Figure \ref{fig:algorithm}B. While $\nu$ is a factor which allows to adjust the tolerance margin, $\lambda$ controls the smoothness of the function. Note that instead of a rectangular function which penalizes deviations in both directions, a simple step function approximated by a single sigmoid could be employed as well since signal dropouts only result in negative differences if the reference value is set appropriately. However, in order to be able to also suppress potential outliers with high signal, for example due to noise or motion, the presented definition of $f(\cdot)$ was used in this work.

The proposed algorithm heavily relies on the determination of a robust reference value $m_i$ which is supposed to represent the true, uncorrupted signal intensity. Although the median is robust against outliers, it can become biased if the majority of repetitions suffers from signal dropouts in a certain image region. Considering the left hepatic lobe at high diffusion-weightings, however, this is not an unlikely scenario.

	\subsection{Deep Learning Classifier}

In order to reduce the risk of employing a potentially biased reference value which could lead to penalizing uncorrupted signal, the median can instead be calculated on a subset of repetitions which is assumed to not contain any signal dropouts. For this purpose, we propose to train a Deep Learning classifier (DLC) to predict whether signal dropout appears in a given repetition.

\begin{figure}[tb]
	\centering
	\includegraphics[width=\textwidth]{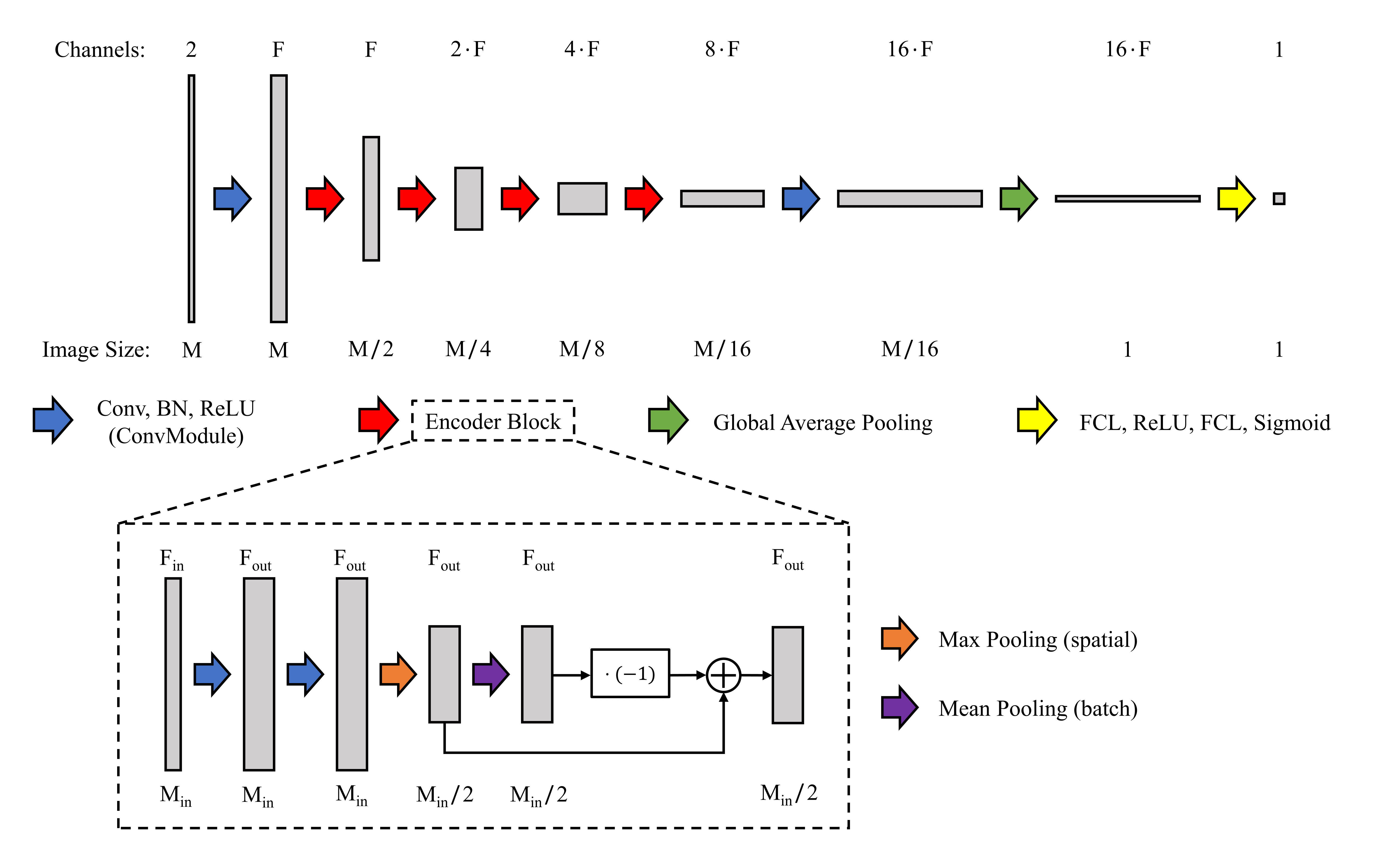}
	\caption{Network architecture of the DLC consisting of a convolutional encoding part and a final classification part. The former is composed of four encoder blocks each of which expands the number of feature maps while reducing their spatial extent. The basic building blocks are ConvModules consisting of a convolution layer (Conv), Batch Normalization (BN) and ReLU activation. The first ConvModule expands the number of channels from 2 to $F$. The initial image size $M$ (using one variable for both image dimensions for simplicity) is halved in every encoder block via spatial max pooling. The resulting feature maps then undergo set aggregation by using mean pooling across the batch dimension. The aggregated feature map is subtracted elementwise from the initial set of feature maps after spatial pooling. After passing through all levels of encoding, feature maps are reduced to single scalars via global average pooling before being fed to two fully-connected layers (FCL) interleaved by a ReLU activation. The probabilistic prediction of the network is yielded by the final sigmoid activation.}
	\label{fig:arch}
	\end{figure}

The employed network architecture presented in Figure \ref{fig:arch} can be separated into a convolutional encoder part and the actual classification part producing a single scalar prediction via global average pooling and fully-connected layers. The final module of the network is a sigmoid activation which scales the prediction in a probabilistic range between 0 and 1. During the encoding part, the image dimensions are decreased while feature channels are generally increased. The encoder’s basic building blocks are convolutional modules (ConvModule) which consist of a convolution layer with a kernel size of $3 \times 3$ followed by Batch Normalization and ReLU activation. The initial convolution layer of the network expands the number of feature channels of the input to $F$. In total, four encoding levels are used, each of which consists of two ConvModules followed by spatial max pooling. Further, joint classification of a set of repetitions is performed in order to exploit shared information. Following the \textit{Deep Set} concept \cite{DeepSets}, the set of repetitions is treated as a typical input batch while aggregating features across the batch dimension by channel- and pixel-wise mean pooling at the end of every encoder block. The pooled features are then distributed back onto the set by subtracting them from every element. This way, central requirements, such as the ability to handle different set sizes as well as permutation-equivariance, can be ensured.

	\subsection{Data}
	
Liver DWI was acquired in 29 healthy volunteers on various 1.5 and 3\,T MR scanners (MAGNETOM Altea, Sola, Lumina, Vida, Siemens Healthcare, Erlangen, Germany) using a prototypical single-shot EPI sequence. Further, one patient data set with focal liver lesions, used for evaluation purposes only, was measured with a clinical protocol on a 1.5\,T scanner (MAGNETOM Aera, Siemens Healthcare, Erlangen, Germany) at the University Hospital Erlangen. Scans were conducted with written informed consent and IRB approval. Relevant acquisition parameters for both protocols are presented in Supporting Information Table \ref{stab:acqparam}. DWI employed monopolar diffusion gradients waveforms and two $b$-values (50 and 800\,s/mm$^2$). Further, all measurements were performed in free-breathing and with parallel acceleration ($R=2$). Images were reconstructed using GRAPPA \cite{GRAPPA} and subsequently coil-combined.

Volunteer data was split on a subject level into training (21 volunteers), validation (4), and test sets (4). Given that 30-35 axial slices were collected per volunteer, training, validation, and test sets comprised 695, 135 and 135 image slices, respectively. In total, 6,065 and 18,210 single image repetitions were available at the lower and higher $b$-value, respectively. Based on the occurrence of signal dropouts, all repetitions of the higher $b$-value were manually labeled as either ``clean" (negative) or ``corrupt" (positive). Data was split into two disjoint sets of approximately the same size and annotated by two non-clinical MR experts, respectively. The patient data set was not annotated as it was treated as a prospective use case.

	\subsection{Implementation and Training}

If not stated otherwise, the AWA used the following hyperparameter configurations: $P=5$, $\nu=1$ and $\lambda=5$. The patch-wise estimation of mean and standard deviation can be efficiently implemented using convolutions with a mean filter kernel. In order to benefit from GPU support, both the AWA and the DLC were implemented using the \textit{Pytorch} framework \cite{Pytorch}. 

Concerning the DLC, the initial number of feature channels was set to $F=16$. Classification was performed on the repetitions acquired at a $b$-value of 800\,s/mm$^2$ only. However, each repetition was concatenated with its respective low $b$-value image (average of all repetitions) resulting in a two-channel input. As the low $b$-value images typically were not affected by signal dropouts, the additionally provided reference information was expected to facilitate the classification task. The inputs were pre-processed by normalization with the 98th percentile value of the higher $b$-value image. Further, data augmentation was performed during training by random rotation (multiples of 90°) and flipping. The objective of training was to minimize the binary cross entropy between the network prediction and the ground-truth label (0: negative, 1: positive). In order to account for the imbalance of the training data set which contained approximately three times more negative than positive repetitions, samples with a positive ground-truth label were weighted by a factor of 3 within the loss function. Optimization was performed by stochastic gradient descent with a learning rate of 10$^{-4}$ which was halved every 20 epochs. Training was stopped when the validation loss did not decrease for 20 epochs.

	\subsection{Evaluation}
	
	\subsubsection{Classification}

The classification performance of the DLC was evaluated by analyzing the receiver operating characteristic (ROC) curve on the test set for different binary thresholds and computing the area under the ROC curve (AUC). Based on the choice of a particular threshold value, classification metrics such as accuracy, sensitivity, specificity, and precision were calculated as well. For this purpose, the threshold maximizing the difference between the true positive rate (TPR) and the false positive rate (FPR) was chosen. In order to conduct a fair evaluation and avoid optimizing on the test set, the ROC curve on the validation set was used for the task of threshold selection.

	\subsubsection{Qualitative Evaluation}

The proposed method of using the AWA with DLC-support in calculating patch-wise reference values will be referred to as \textit{Deep Learning-guided Adaptive Weighted Averaging} (DLAWA) in the following. For a representative set of repetitions from the test split, weight maps as well as the resulting average image were inspected. In order to validate the proposed approach, it was compared to three other methods: 1) uniform averaging of all available repetitions, 2) the conventional AWA without classification support and 3) uniform averaging of the repetitions which are marked as uncorrupted by the classifier, as proposed in \cite{Tamada}. The latter will be referred to as \textit{Classify and Discard} (C\&D). In addition to the DW images, ADC maps were analyzed as well. Given $N$ repetitions in total for a slice, ground-truth data was generated by uniform averaging of the subset of uncorrupted repetitions of cardinality $N_0$. Accordingly, the input to the different methods was generated by taking a random subset of repetitions of the same cardinality $N_0$ which could contain corrupted repetitions as well.
	
	\subsubsection{ADC Quantification}

Within the test set, a region-of-interest (ROI) was placed into the left liver lobe of every slice that contained it (37 out of 135 slices). ADC maps were calculated for this subset of slices based on the images produced by DLAWA and the three comparison methods with ground-truth and input data generated as described in the previous paragraph. In order to make a broader evaluation, 15 distinct runs were performed across which the composition of input repetitions randomly varied for a given slice. Further, the performance of the methods was analyzed with respect to different ratios of corrupted repetitions within an input set. Therefore, the dropout ratio of a set was defined as $100\,\% \cdot (1-\frac{N_0}{N})$, where $N_0$ and $N$ again denote the cardinalities of the uncorrupted subset and total set of repetitions, respectively. The mean ADC within ROIs was calculated and assigned to bins of dropout ratios of 10\,\% width. Within each bin, ADC values were averaged across slices and runs for every method.

	\subsubsection{Noise Analysis}
	
One important aspect of the proposed method is its potential SNR-saving property. It was compared against the purely classification-based approach by analyzing respective noise maps. For the purpose of deriving noise analytically, we relied on the assumption of Gaussian noise with zero mean. Although Rician distributions are known to model noise in magnitude MR images more appropriately, the true amount of noise was expected to scale well with a Gaussian model. Further, it was assumed that the noise standard deviation $\sigma_i$ was equivalent in all repetitions for a given pixel location $i$ (not to be confused with the estimated standard deviation $\bs{\sigma}_i$ within patches used in section \ref{sec:AWA}). Hence, the standard deviation of a weighted sum of $N$ normally distributed random variables with zero mean scales as $\sqrt{\sum_{n=1}^N a_{i,n}^2} \cdot \sigma_i$, where $a_{i,n}$ are the respective weights. Accordingly, in the case of uniform averaging where $a_{i,n}=\frac{1}{N}$, the resulting standard deviation of the noise is
\begin{equation}
	\sigma_i^{\text{uni}} = \sqrt{\frac{N}{N^2}} \cdot \sigma_i = \frac{1}{\sqrt{N}} \cdot \sigma_i \;.
\end{equation}
Noise maps were scaled relatively to uniform averaging which yields the minimum achievable noise as all repetitions contribute equally. Hence, the relative noise amplification of C\&D and DLAWA could be computed as follows:
\begin{align}
	\sigma_{\text{rel},i}^{\text{C\&D}} &= \frac{\sigma_i}{\sqrt{N_0'}} \cdot \frac{1}{\sigma_i^{\text{uni}}} = \sqrt{\frac{N}{N_0'}}\\[1em]
	\sigma_{\text{rel},i}^{\text{DLAWA}} &= \frac{\sqrt{\sum_{n=1}^N \widetilde{w}_{i,n}^2} \sigma_i}{\sigma_i^{\text{uni}}} = \sqrt{N \cdot \sum_{n=1}^N \widetilde{w}_{i,n}^2}
\end{align}
where $N_0'$  denotes the number of uncorrupted repetitions as predicted by the classifier and $\widetilde{w}_{i,n}$  are the normalized weights for DLAWA as calculated in Equation \ref{eq:weights}. Similar to the ADC analysis described previously, noise characteristics of both methods were assessed quantitatively for different dropout ratios across 15 different runs. However, in contrast to the ADC which was calculated on ROIs in a subset of the test set which contained the left liver lobe, noise was quantified by computing the average of entire noise maps across the complete test set.

	\subsubsection{Hyperparameter Influence}

In order to test the controllability of DLAWA, the effect of the algorithm’s hyperparameters on the results were investigated. While $\nu$ was fixed to 1 meaning that the tolerance was set to one standard deviation as estimated on the patches, three different configurations of $\lambda$ (1, 5, 25) were compared. Two repetitions from a representative set were selected ‒ one corrupted and one uncorrupted by signal dropouts ‒ and weight maps resulting from the different hyperparameter configurations were visually analyzed along with resulting averages, ADC and noise maps.

	\subsubsection{Clinical Case}

The ability of DLAWA to deal with signal dropouts in images with pathologies was tested on a slice from the patient data set which exhibited focal liver lesions. Again, DLAWA was compared to uniform averaging, AWA and C\&D in terms of quality of DW images and respective ADC maps. As no labels were available for the patient data, a ground-truth resulting from averaging of uncorrupted images was not part of the comparison.

	\section{Results} \label{sec:results}
	
	\subsection{Classification}
	The DLC achieved an AUC score of 0.87 on the test set with the corresponding ROC curve presented in Supporting Information Figure \ref{sfig:roc}. The threshold value to maximize the difference between TPR and FPR on the validation set was found to be 0.68. Based on this, the DLC achieved an accuracy of 0.82, a sensitivity (TPR) of 0.80, a specificity ($1-\text{FPR}$) of 0.83 and a precision of 0.42 on the test set.

	\subsection{Qualitative Evaluation}

\begin{figure}[tb]
	\centering
	\includegraphics[width=\textwidth]{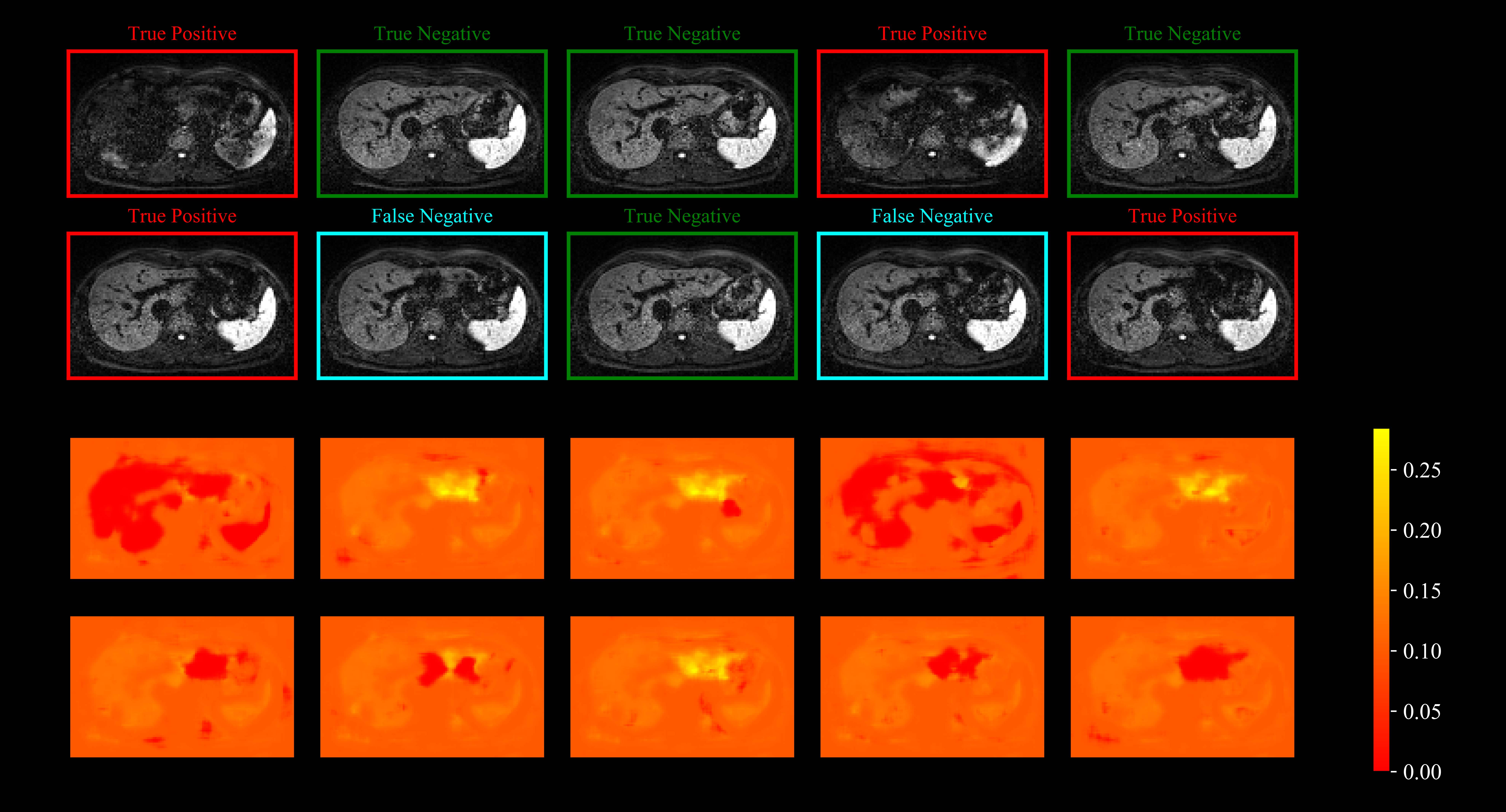}
	\caption{Upper rows: ten repetitions of a liver slice acquired at a $b$-value of 800\,s/mm$^2$ with six repetitions being affected by signal dropouts of varying severity. Predictions of the DLC for the individual repetitions are indicated above images with color-coding. While all four negative repetitions were classified correctly, two of the corrupted repetitions had been misclassified as negative. Bottom rows: corresponding weight maps estimated by DLAWA. It becomes apparent that regions of lower weight closely correspond with locations of signal dropouts. Accordingly, the respective image regions are assigned higher weight in uncorrupted repetitions to ensure that weights sum up to one for every pixel.}
	\label{fig:weight_maps}
	\end{figure}

Figure \ref{fig:weight_maps} shows a representative set of repetitions of a liver slice and the corresponding normalized weight maps estimated by DLAWA. In 6 out of 10 repetitions, signal dropouts of different severities can be observed which primarily affect the left lobe but also the spleen and other parts of the liver. Regions with lower values in the weight maps spatially coincide with signal dropouts. Although the DLC misclassified two of the repetitions with moderate signal dropouts as negative, the proposed algorithm is still able to compensate for this by assigning lower weight to corrupted pixels in the respective images.

\begin{figure}[tb]
	\centering
	\includegraphics[width=\textwidth]{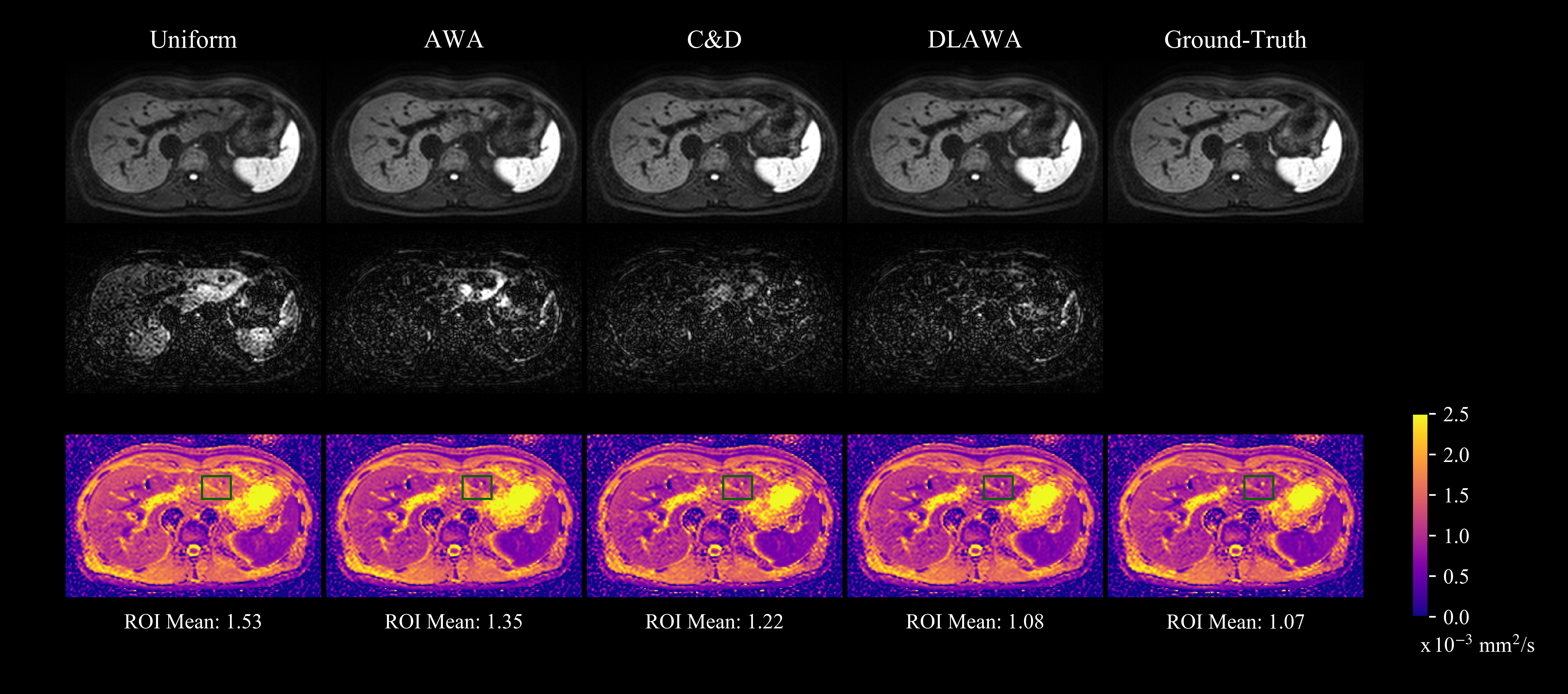}
	\caption{Qualitative evaluation of methods on the set of repetitions shown in Figure \ref{fig:weight_maps}. First row: images produced by the different methods. Second row: difference images with respect to the ground-truth magnified by a factor of 5. Third row: corresponding ADC maps with an ROI (green rectangle) placed into the left liver lobe. Mean ADC within the ROI is indicated below. Conventional uniform averaging leads to signal reduction in the left liver lobe and corresponding overestimation of ADC. The AWA suffers from a biased median at several pixel locations which results in even more pronounced signal attenuation. C\&D yields slightly reduced signal in the left lobe as it is impacted directly by misclassification of the DLC which contribute to the average. The results obtained by applying the previously shown weight maps produced by DLAWA come very close to the ground-truth in terms of qualitative image impression as well as ADC quantification within the ROI.}
	\label{fig:qual_comp}
	\end{figure}

Applying the weight maps accordingly leads to a homogenized liver signal compared to uniform averaging as presented in Figure \ref{fig:qual_comp}. The attenuated signal of the left lobe in the latter case introduces substantial bias into the ADC. Compared to the ground-truth, the ADC is overestimated by 43\,\% inside the ROI which can be reduced to a bias of approximately 1\,\% using DLAWA. The plain AWA alleviates the signal loss in large parts of the image but fails to do so at some locations within the left lobe where it even appears that signal loss is aggravated. C\&D is able to mitigate signal attenuation but not to the level of DLAWA since two false negatives contribute to the average.

	\subsection{ADC Quantification}

\begin{figure}[tb]
	\centering
	\includegraphics[width=\textwidth]{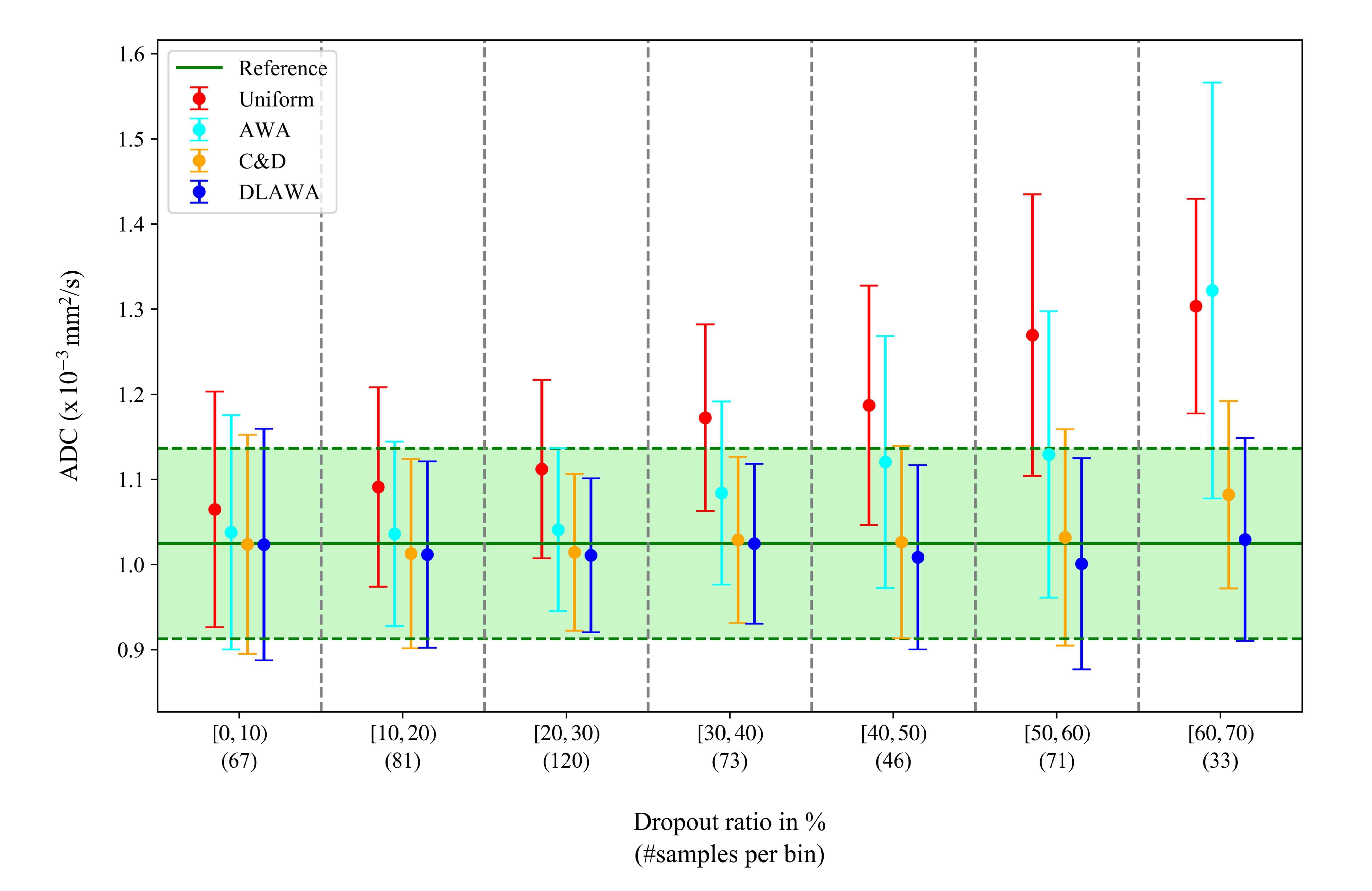}
	\caption{Quantitative analysis of ADC in the left liver lobe across different ranges of dropout ratios. As expected, ADC bias grows with increasing dropout ratio in the case of uniform averaging. The same applies to AWA with ADC staying relatively constant for dropout ratios of up to 30\,\% only. Both C\&D and DLAWA produce ADC values which do not deviate substantially from the range of reference values obtained from uniform averages of exclusively uncorrupted repetitions.}
	\label{fig:adc_quant}
	\end{figure}

The quantitative ADC evaluation presented in Figure \ref{fig:adc_quant} confirms the observations of the qualitative comparison. The reference ADC of the left liver lobe calculated on averages of uncorrupted repetitions was $(1.02 \pm 0.11)\cdot 10^{-3}$ mm$^2$/s. Overestimation of ADC in uniform averaging was approximately proportional to the dropout ratio. The remaining methods performed similarly well for dropout ratios of up to 30\,\%. For ratios of 50\,\% and higher, the AWA was not able to prevent substantial ADC bias. In contrast, both DLAWA and C\&D produced ADC values that generally were in agreement with the reference across dropout ratios.

	\subsection{Noise Analysis}

\begin{figure}[tb]
	\centering
	\includegraphics[width=.8\textwidth]{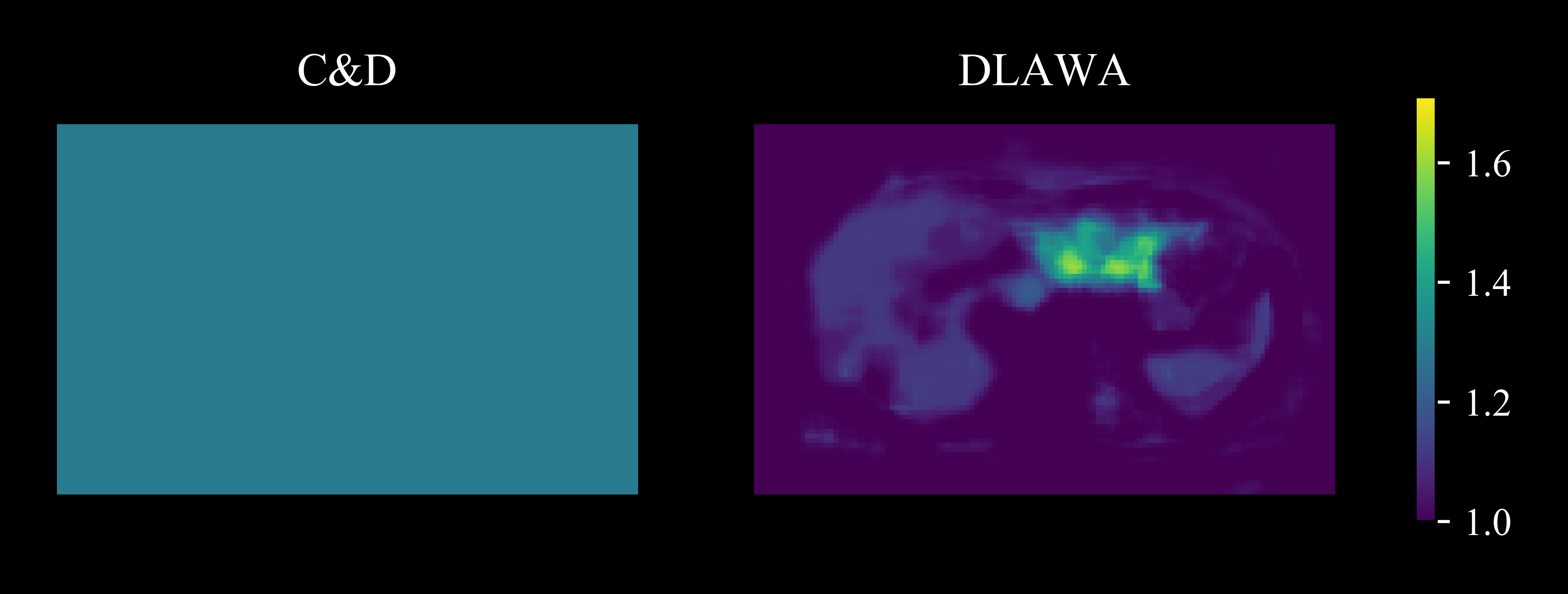}
	\caption{Comparison of noise maps computed for the example presented in Figure \ref{fig:weight_maps}. Both maps are normalized with respect to uniform averaging in order to visualize the relative noise enhancement. As entire image repetitions are excluded from the final image computation in C\&D, noise is amplified globally compared to uniform averaging. Note that this amplification would be even more profound if no false negatives were predicted. As DLAWA performs a locally aware weighted average, noise is kept to a minimum in large parts of the image and is enhanced only in regions where it is necessary to suppress signal dropout artifacts.}
	\label{fig:noise_maps}
	\end{figure}

\begin{figure}[tb]
	\centering
	\includegraphics[width=\textwidth]{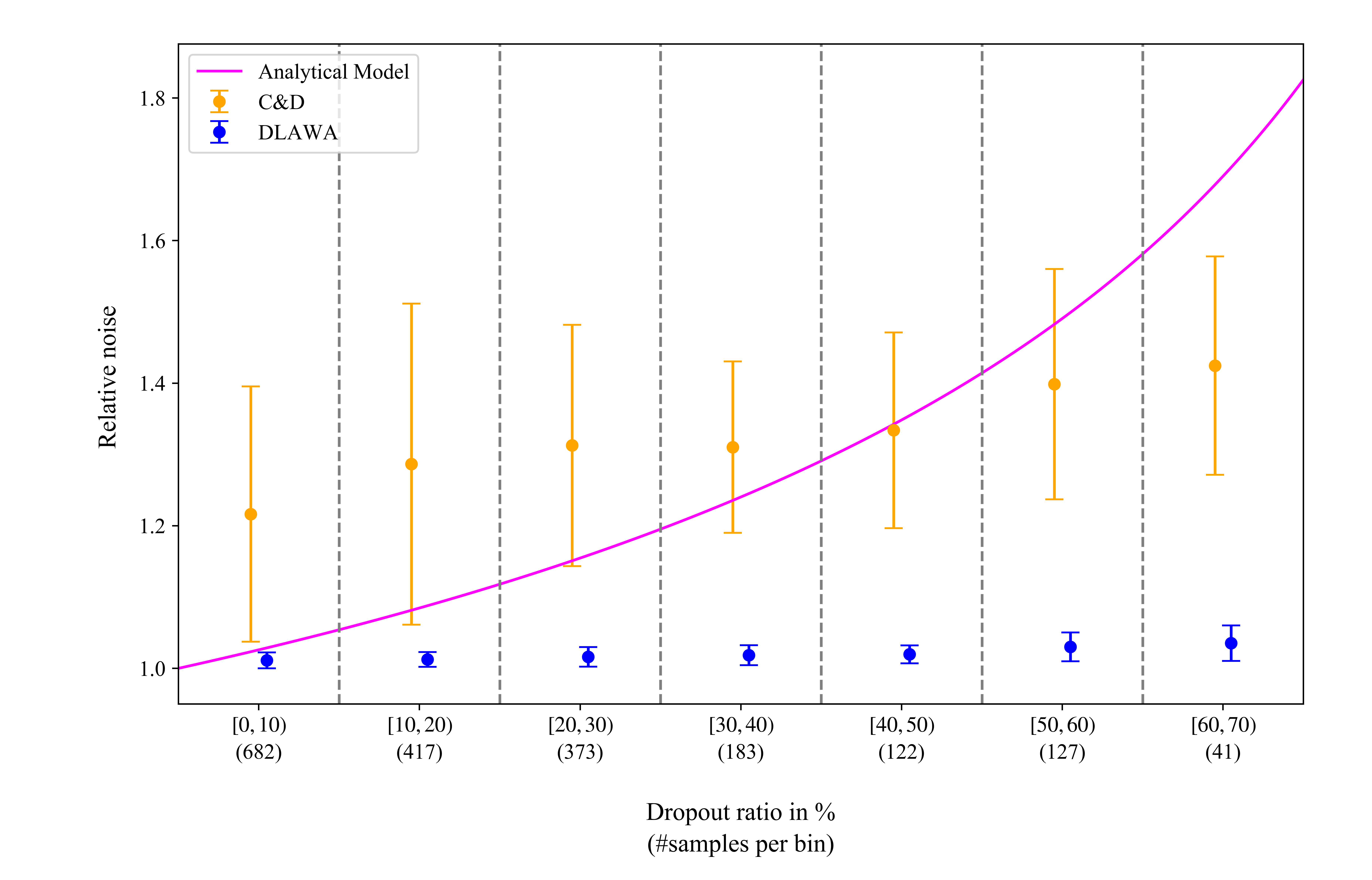}
	\caption{Quantitative evaluation of mean relative noise enhancement in images produced by C\&D and DLAWA across different intervals of dropout ratios. The ideal analytical model for C\&D shown in magenta is based on the assumption that the classifier predicts the correct percentage of negatives in a set of repetitions. In general, DLAWA leads to significantly less global noise increase compared to C\&D. Because the classifier has a relatively low precision and therefore produces several false positives, C\&D suffers from immoderate noise amplification at low dropout ratios. Because there also seem to be false negatives in regimes of high dropout ratios, the results of C\&D do not follow the model curve.}
	\label{fig:noise_quant}
	\end{figure}

Figure \ref{fig:noise_maps} shows the noise maps of C\&D and DLAWA which were computed based on the example in Figure \ref{fig:weight_maps}. Since 4 out of 10 repetitions are entirely discarded using C\&D, the noise globally increases by a factor of $\sqrt{\frac{10}{6}} \approx 1.29$ compared to uniform averaging. Note that if all 6 corrupted repetitions had been detected, the relative noise amplification would be higher. In contrast, DLAWA increases noise only locally in regions which are affected by dropouts while maintaining noise levels close to uniform averaging in uncorrupted image regions. The noise quantification on the entire test set presented in Figure \ref{fig:noise_quant} reveals that DLAWA yielded substantially lower noise than C\&D across all dropout ratios. As the classifier is not perfect, the results of C\&D do not follow the model curve which represents the relative noise enhancement if the correct number of corrupted repetitions had been predicted.

	\subsection{Hyperparameter Influence}
	
\begin{figure}[tb]
	\centering
	\includegraphics[width=\textwidth]{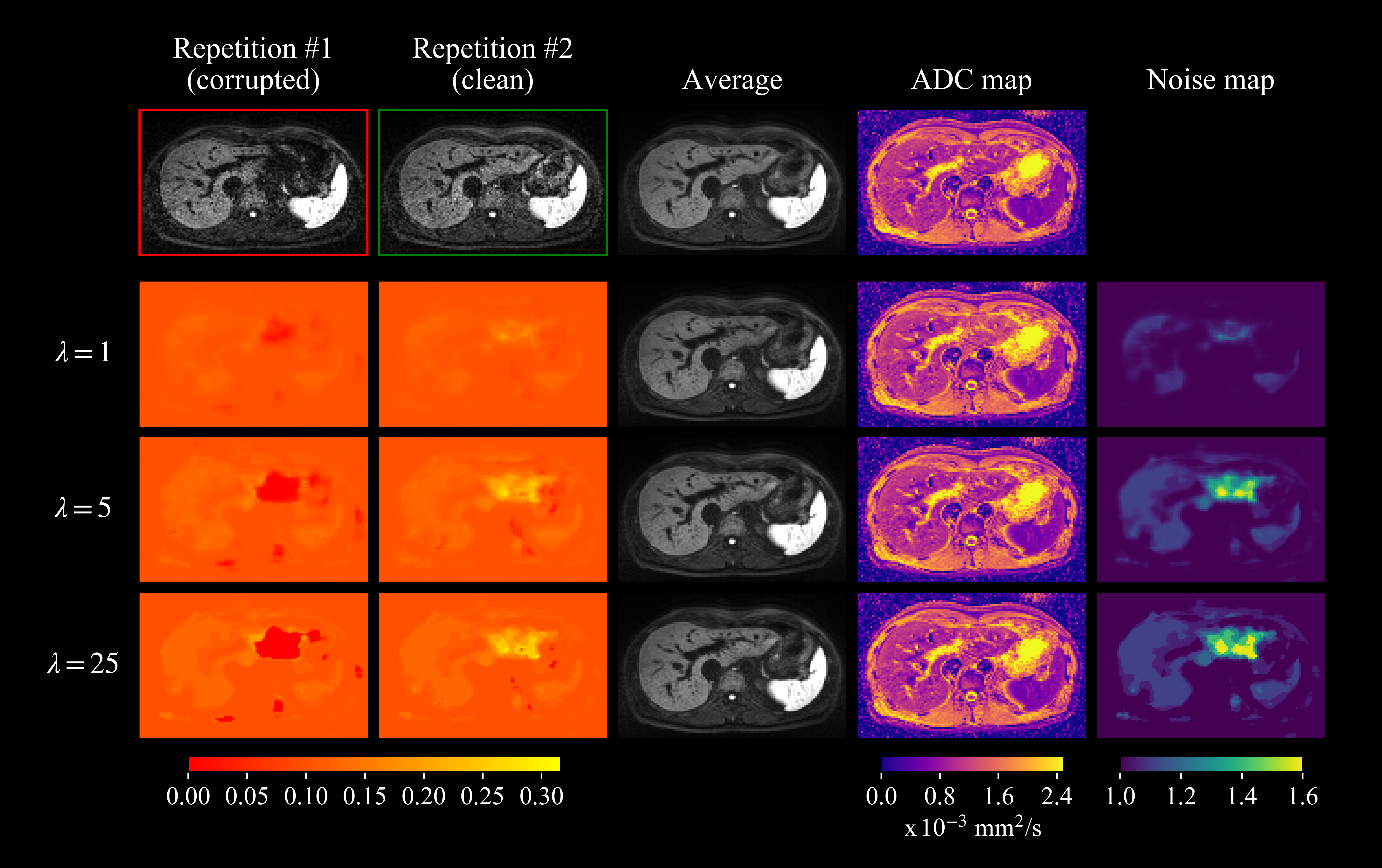}
	\caption{Qualitative evaluation of hyperparameter influence on DLAWA results. First two columns: weight maps for two selected repetitions of the entire set shown in Figure \ref{fig:weight_maps}. With growing $\lambda$ the differences in weights between corrupted and uncorrupted image regions become more profound. As a result, signal dropout compensation in corresponding averages becomes more effective (third column) which also reduces ADC bias (fourth column). Simultaneously, noise amplification in relation to uniform averaging is the highest at large $\lambda$ (fifth column). An intermediate value for $\lambda$ can be used in order to achieve a trade-off between dropout suppression and noise penalties. The uppermost image in the third and foruth column, respectively, displays the ground-truth generated from exclusively uncorrupted image repetitions.}
	\label{fig:hyperparams}
	\end{figure}

Results of the qualitative hyperparameter analysis are provided in Figure \ref{fig:hyperparams}. For small $\lambda$, DLAWA produces smooth weight maps which achieve insufficient dropout suppression as apparent in the attenuated left liver lobe of the average and the corresponding ADC bias. However, relative noise enhancement is minimal. At $\lambda=25$, the difference in weights between uncorrupted and corrupted image regions is substantially higher. This enables better recovery of DW signal and reduction of ADC bias in the left lobe while introducing more SNR penalty. The latter can be reduced when choosing an intermediate value of $\lambda=5$ which constitutes a reasonable trade-off as it still provides satisfactory dropout suppression.

	\subsection{Clinical Case}

\begin{figure}[tb]
	\centering
	\includegraphics[width=\textwidth]{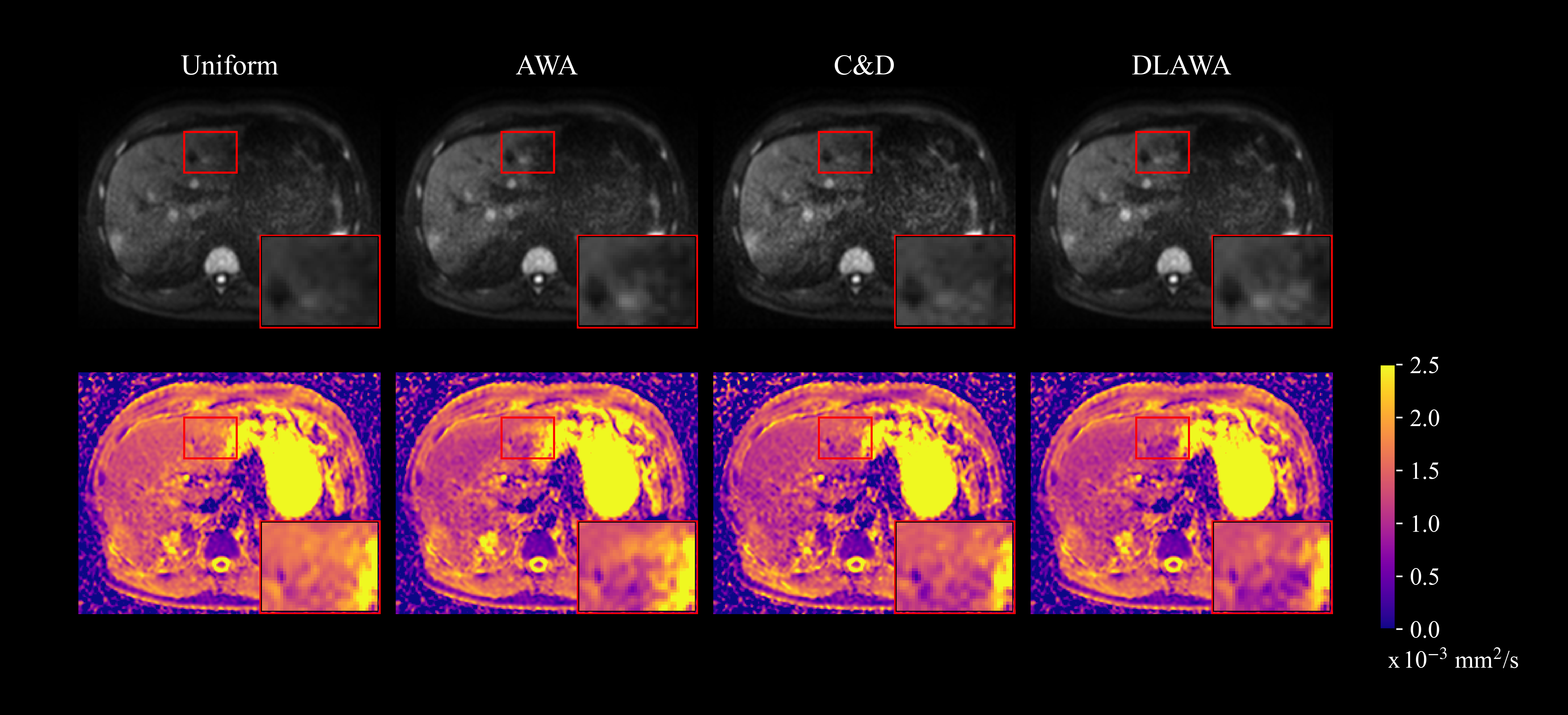}
	\caption{Evaluation on a patient case with several focal liver lesions. First row: images produced by the different methods. Second row: corresponding ADC maps. The lesion located in the left lobe appears diffuse and low in signal in the uniform average as highlighted in the zoomed-in ROI. Compared to the other methods, DLAWA produces an image in which the respective lesion appears more distinct. Especially with respect to C\&D, the lesion is substantially less obscured by noise.}
	\label{fig:patient}
	\end{figure}

As displayed in Figure \ref{fig:patient}, DLAWA is able to recover signal in the left liver lobe compared to uniform averaging. It manages to do that more effectively than AWA and C\&D, the latter of which exhibits evidently lower SNR due to the exclusion of 6 out of 12 repetitions. Moreover, the zoomed-in region within the left liver lobe shows that DLAWA enables better visualization and delineation of a focal liver lesion and adjacent hyperintense structures. The single repetitions along with the respective classifier predictions and weight maps produced by DLAWA are presented in Supporting Information Figure \ref{sfig:patient_weights}.

	\section{Discussion} \label{sec:discussion}

The results of the conducted experiments demonstrate the value of the proposed method for signal dropout compensation in abdominal DWI. The spatially resolved weight maps produced by DLAWA appear reasonable as areas of low weight accurately overlay with signal dropouts in the corresponding image repetitions. Consequently, dropouts can be suppressed locally which prevents corresponding ADC bias further down-stream. The evaluation on the liver slice from the patient data set underlines the potential clinical benefits of the method as it is able to recover well-delineated focal lesions which appear less clear in the uniform average as a consequence of signal dropouts in individual repetitions.

The conventional AWA, which lacks the guidance of a classifier, is limited in cases with high prevalence of signal dropouts. If the majority of repetitions is affected, the median becomes biased at certain pixel locations which leads to down-weighting of uncorrupted signal. This is confirmed by the quantitative analysis of ADC values in Figure \ref{fig:adc_quant} as AWA produces even higher ADC bias than uniform averaging at dropout ratios of 60 to 70\,\%. In terms of ADC quantification, both DLAWA and C\&D perform comparably well across dropout ratios. However, the latter has the disadvantage of low SNR efficiency as entire image repetitions are discarded. Consequently, the global noise level in the averaged images is substantially higher than for DLAWA which sacrifices SNR only locally in regions of signal dropouts. The quantitative analysis of noise across dropout ratios presented in Figure \ref{fig:noise_quant} reveals that the results of C\&D do not follow the analytical model curve, which visualizes the relative amount of noise in the case of perfect classification, but rather stay comparatively constant across dropout ratios. The relatively low precision of 0.42 is further evidence that the classifier produces a certain number of false positives leading to excessively increased noise at lower dropout ratios. On the other hand, noise is less pronounced compared to the model curve for high dropout ratios implying a certain number of false negatives as well. Note that these relationships could be influenced by adjusting the binary threshold for the classifier. For instance, by setting it higher, less false positives would occur resulting in lower SNR penalties at lower dropout ratios. However, this would come at the cost of an increased prevalence of false negatives which may impede effective dropout suppression.

While misclassifications have an immediate impact on the results of C\&D, DLAWA is more robust against them. As shown in the example of Figure \ref{fig:weight_maps}, reasonable weight maps are estimated although two false negatives are present. Hence, as long as there are more true than false negatives in a set, the median is expected to be unbiased. On the other hand, false positives can still be assigned full weight if the differences from the median lie within the tolerance. In summary, compared to C\&D the advantages of the proposed approach are two-fold: 1) less global noise penalties and 2) less sensitivity to misclassifications.

In general, there exists a trade-off between the effective suppression of signal dropouts and SNR which is dependent on the entropy of the estimated weights. The hyperparameters used in DLAWA provide a transparent way of controlling this trade-off. By fixing $\nu$ as a factor between 1 and 2, small $\lambda$ favor more homogenous weights (low entropy) while large $\lambda$ result in weights with higher spread (high entropy). The higher the entropy is with respect to the weights, the higher becomes the effectiveness of the method in suppressing signal dropouts, however, at the cost of increasing noise. In the opposite extreme case of $\lambda=0$, DLAWA reduces to uniform averaging yielding no dropout suppression but minimum possible noise. In this work, a medium value of $\lambda=5$ was employed in order to achieve a reasonable compromise between dropout suppression and SNR.

Similarly to related works following a post-processing approach to eliminate signal dropouts, this method relies on the assumption that at least one image in the set of repetitions is free of dropout artifacts. In cases where this assumption is violated, DLAWA will not be able to compensate for signal loss. Nevertheless, the value of the proposed method primarily shows in scenarios of high dropout ratios which necessitate the classifier-based filtering of repetitions for median computation. In order to demonstrate the frequency of signal dropouts in real data, dropout ratios were analyzed and visualized in a histogram (see Supporting Information Figure \ref{sfig:hists}). This was done once for the entire data set (training, validation, test) also containing slices above and below the liver, and once on the subset of slices from the test set containing the left liver lobe. Across the entire data set, the percentage of cases with dropout ratios of over 50\,\% was 13.8\,\%. However, in the latter case this percentage amounted to 21.6\,\% which underlines the sensitivity of the left liver lobe to intra-voxel dephasing but also demonstrates the necessity for a method that addresses the problem of signal dropouts with high prevalence.

This work focuses on the effects of motion occurring within an EPI shot (intra-repetition motion) primarily caused by the cardiovascular system. However, respiratory and bulk motion between repetitions (inter-repetition motion) can occur as well, especially in free-breathing acquisitions, leading to misalignment of individual repetitions and motion blurring of averages. It is demonstrated in Supporting Information Figure \ref{sfig:motion_weights} and \ref{sfig:motion_comp} that the AWA can be used to compensate for these kinds of motion differences as well by down-weighting image pixels which are inconsistent with a target motion state. Hence, the patch-wise reference was not defined as the median of a subset of repetitions found to be uncorrupted by a classifier. Instead, the median was computed on the repetitions belonging to the end-expiration motion state. Note that this can be combined with the proposed form of DLAWA to simultaneously compensate for both intra- and inter-repetitions motion, however, this was considered beyond the scope of this manuscript.

One limitation of this work is the quality of annotation used for training the classifier. First, labels were assigned by two non-clinical MR experts whose proficiency in evaluating signal dropouts was limited. Secondly, because annotations were performed for extended periods of time, results were subject to both inter- and intra-observer variability. Expert-level annotations by radiologist would help in improving the quality of the data and the performance of the classifier. Still, the results of this work underline that the proposed method is robust against misclassifications to some extent, achieving reasonable results even without costly expert annotations. With respect to future work, it would be worth investigating how DLAWA can impact the accuracy and reproducibility of diffusion quantification in a clinical setting.

	\section{Conclusion} \label{sec:conclusion}

This work demonstrates the utility of a locally adaptive weighted averaging of image repetitions to correct for signal dropouts caused by cardiac motion in abdominal DWI. Experiments conducted in this work show that signal dropouts can be effectively suppressed by DLAWA. Compared to uniform averaging, signal attenuation within the liver and resulting ADC overestimation can be substantially mitigated. Evaluations performed on patient data also imply the usefulness of the proposed method in restoring signal of diagnostically relevant structures, such as lesions, in areas of the liver prone to signal dropouts. Further, DLAWA achieves a high degree of dropout suppression with considerably less noise penalties than other retrospective methods used for this purpose. Since DLAWA does not require modifications to the acquisition and allows to control the trade-off between dropout suppression and SNR with a few hyperparameters, it constitutes an easy-to-use method for effectively addressing the problem of signal dropouts in abdominal DWI.

	\section*{Acknowledgement}
We would like to thank the Digital Health Innovation Platform (d.hip) for supporting this work. Also, we appreciate the efforts of Majd Hlou in helping to acquire and annotate the data that was used in this work. Further, we express gratitude to Frederik Laun for making it possible to test our method on clinical data.

	\section*{Conflict of Interest}
	Fasil Gadjimuradov receives PhD funding from Siemens Healthcare GmbH. Thomas Benkert and Marcel Dominik Nickel are employees of Siemens Healthcare GmbH. 

	\footnotesize

	\clearpage
	\section*{Supporting Information}
	\captionsetup[figure]{name=Supporting Information Figure}
	\setcounter{figure}{0}
	\begin{figure}[H]
		\centering
		\includegraphics[width=.8\textwidth]{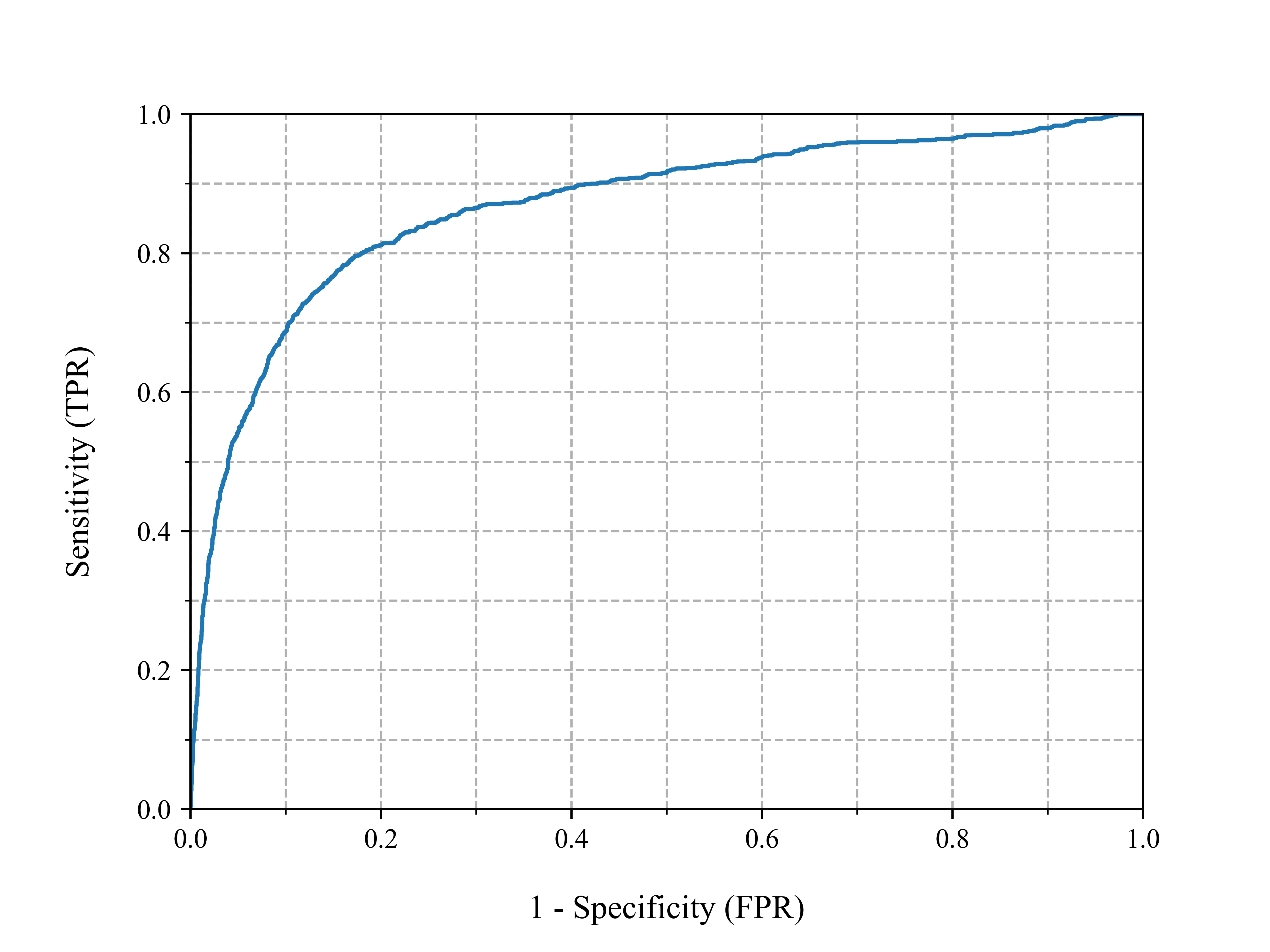}
		\caption{ROC curve of the DLC on the test set with an AUC of 0.87.}
		\label{sfig:roc}
	\end{figure}

	\begin{figure}[H]
		\centering
		\includegraphics[width=\textwidth]{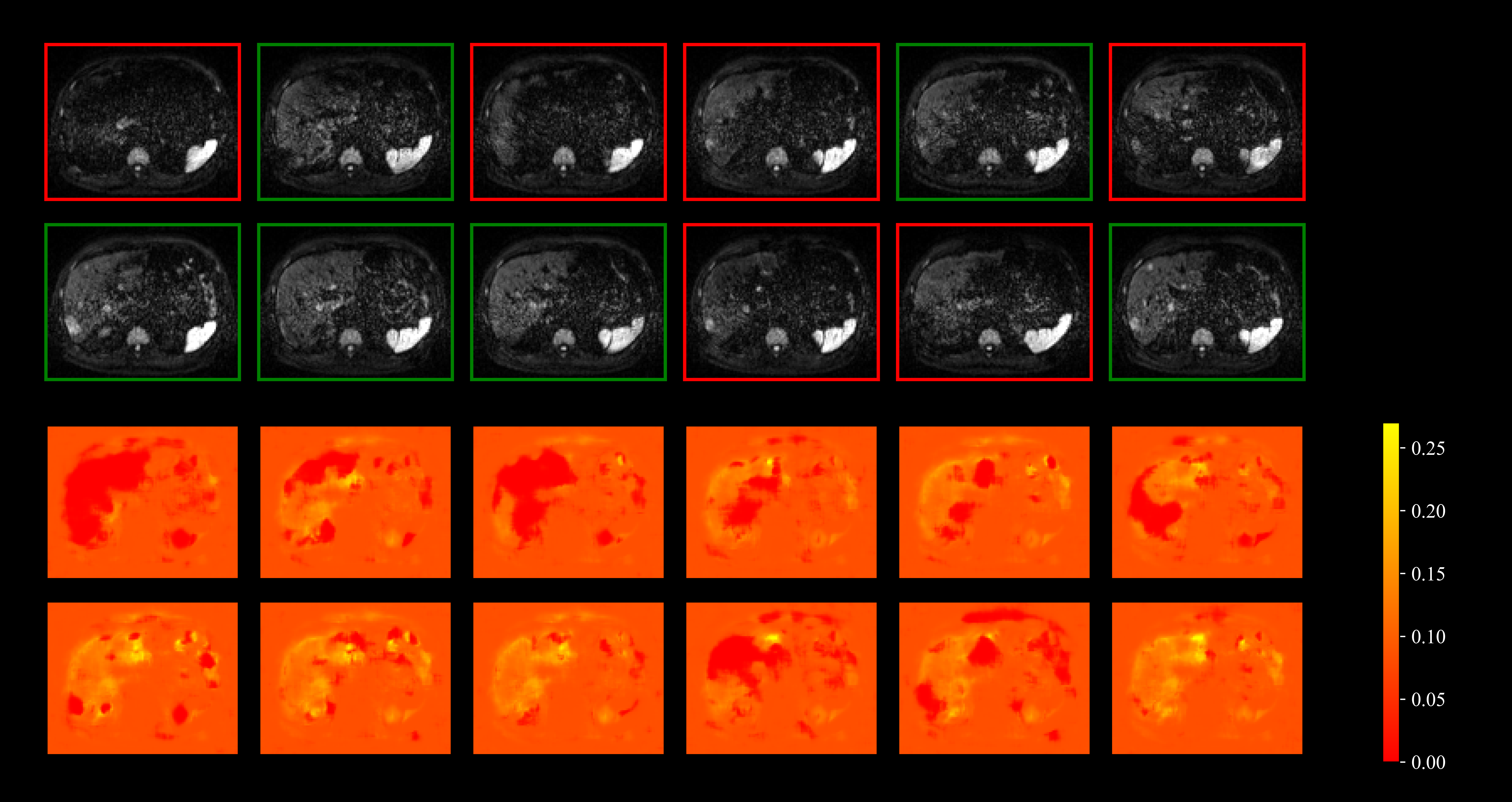}
		\caption{Upper rows: twelve repetitions of a patient liver slice acquired at a $b$-value of 800\,s/mm$^2$ corresponding to the image shown in Figure \ref{fig:patient}. Predictions of the DLC for the individual repetitions are indicated by the color of the frames where red and green define positive and negative predictions, respectively. Note that no ground-truth labels were assigned in this case. Bottom rows: corresponding weight maps estimated by DLAWA.}
		\label{sfig:patient_weights}
	\end{figure}
	
	\begin{figure}[H]
		\centering
		\includegraphics[width=\textwidth]{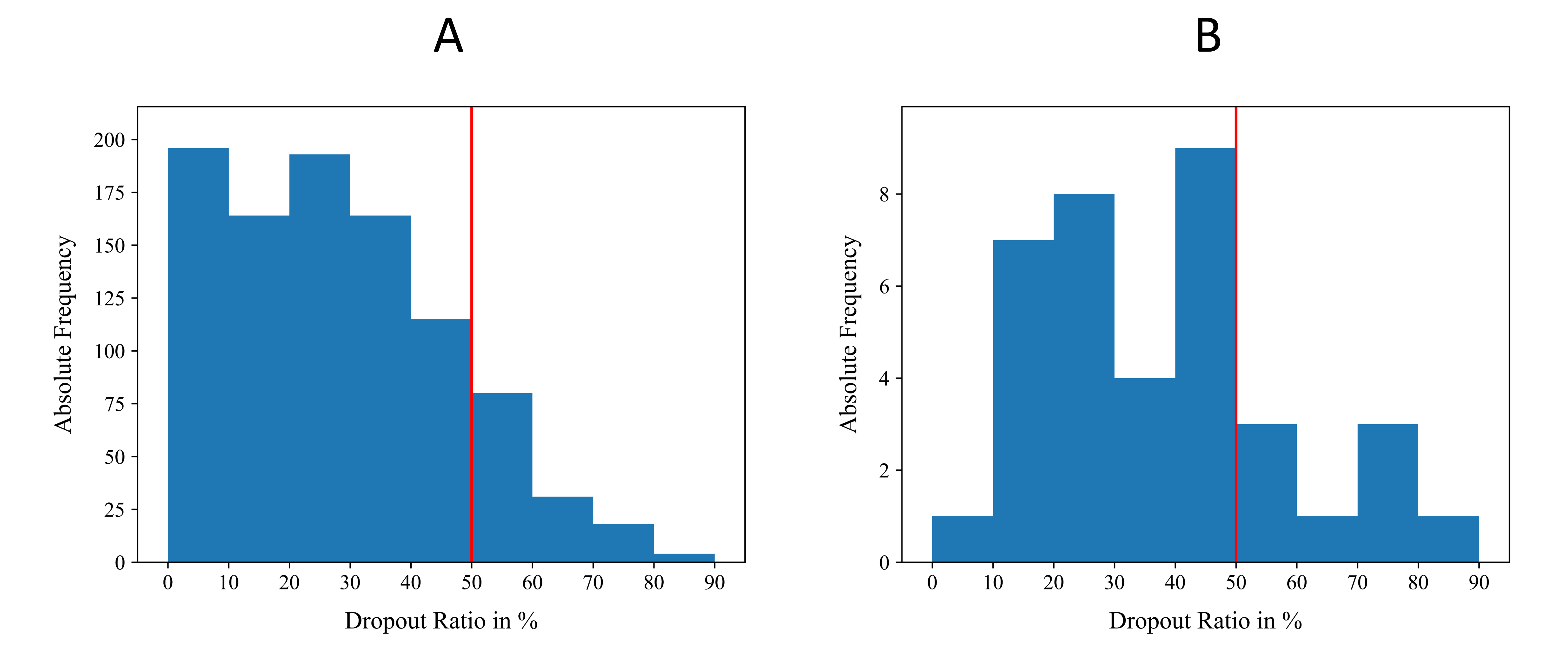}
		\caption{Histogram of dropout ratios (10\,\% bins) across the entire available data set (training, validation, test) \textbf{(A)} and the subset of the test split containing the left liver lobe \textbf{(B)}. The percentage of slices with majorities of corrupted repetitions was 13.8\,\% and 21.6\,\% for the former and latter case, respectively.}
		\label{sfig:hists}
	\end{figure}

	\begin{figure}[H]
		\centering
		\includegraphics[width=\textwidth]{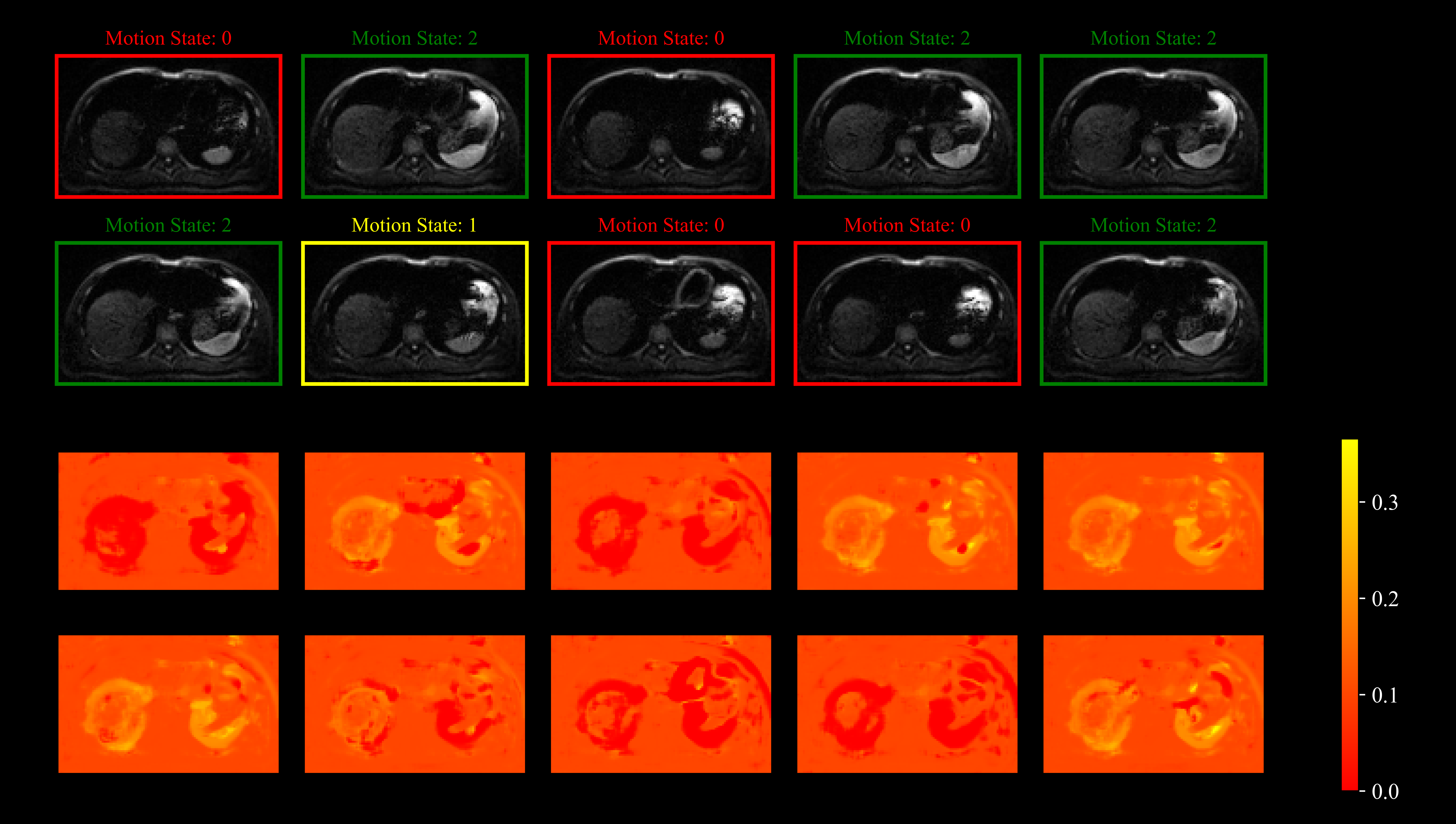}
		\caption{Upper rows: ten repetitions of an axial slice displaying the superior part of the right liver lobe which were acquired at a $b$-value of 800\,s/mm$^2$ and retrospectively gated into three respiratory motion states (0, 1, 2) ordered from end-inspiration to end-expiration (ignoring hysteresis effects). At end-inspiration, the expanded lung pushes the liver into the inferior direction resulting in the acquisition of the most superior part of the right lobe. In contrast, at end-expiration, the lung compression allows the lobe to move into the superior direction such that a more inferior part of the right lobe is being imaged which is larger in axial area. Lower rows: corresponding weight maps obtained by the AWA where the patch-wise median was computed on the repetitions belonging to the end-expiration motion state only. Consequently, the periphery of the lobe in motion states other than end-expiration is down-weighted as it is primarily composed of background due to the lobe being smaller in area in those motion states.}
		\label{sfig:motion_weights}
	\end{figure}

	\begin{figure}[H]
		\centering
		\includegraphics[width=\textwidth]{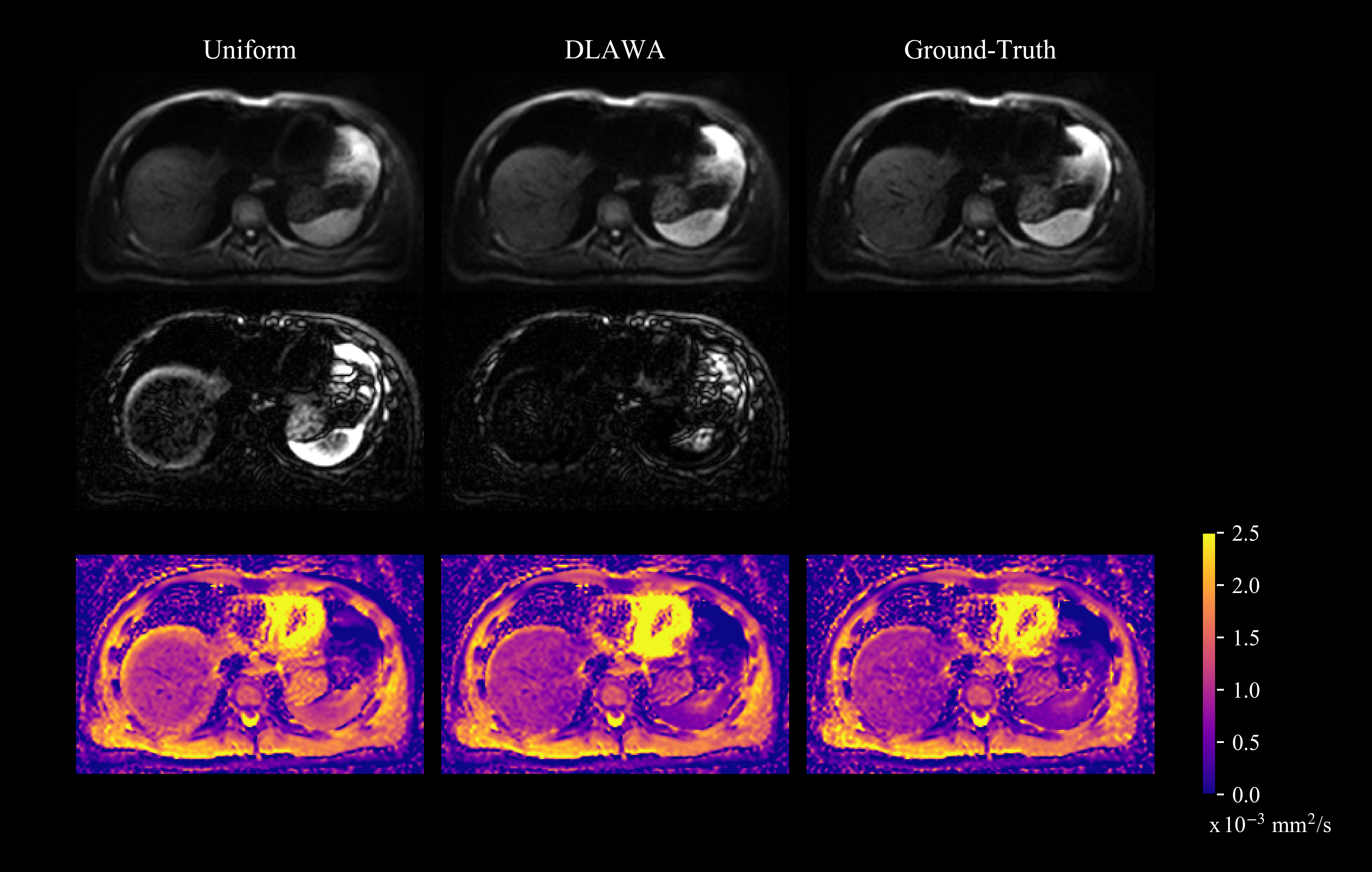}
		\caption{Uniform averaging across all motion states leads to a ring of lower signal intensity around the right liver lobe in the axial view. This further leads to ADC bias in the corresponding area as shown in the bottom row. In contrast, the modified AWA can effectively compensate for the differences introduced by the undesired motion states as DW signal in the liver as well as the corresponding ADC appear more homogeneous and comparable to the ground-truth generated by uniformly averaging ten repetitions from the target motion state.}
		\label{sfig:motion_comp}
	\end{figure}
	
	\captionsetup[table]{name=Supporting Information Table}
	\setcounter{table}{0}
	\begin{table}[H]
		\centering
		\small	
		\caption{List of relevant acquisition parameters of the volunteer ($n=29$) and patient data set ($n=1$). Note that some parameters varied across the volunteer data set.}
		\newcolumntype{C}[1]{>{\centering\arraybackslash}m{#1}}
		\newcolumntype{L}[1]{>{\raggedright\arraybackslash}m{#1}}
		\begin{tabular}{L{5cm}C{1.25cm}C{1.25cm}C{1.25cm}C{1.25cm}}\toprule
			& \multicolumn{2}{c}{Volunteer data set} & \multicolumn{2}{c}{Patient data set}\\\midrule
			$b$-values [s/mm$^2$] & 50 & 800 & 50 & 800\\
			No.\ of repetitions & $5-10$ & $15-25$ & 4 & 12\\
			B0 [T] & \multicolumn{2}{c}{1.5 \& 3} & \multicolumn{2}{c}{1.5}\\
			TE [ms] & \multicolumn{2}{c}{$52-61$} & \multicolumn{2}{c}{70}\\
			TR [ms] & \multicolumn{2}{c}{$5,500-6,300$} & \multicolumn{2}{c}{12,400}\\
			Matrix size & \multicolumn{2}{c}{$134 \times 108$} & \multicolumn{2}{c}{$128 \times 104$}\\
			Resolution (isotropic) [mm$^2$] & \multicolumn{2}{c}{$2.8-3.2$} & \multicolumn{2}{c}{3.1}\\
			Slice thickness [mm] & \multicolumn{2}{c}{5} & \multicolumn{2}{c}{5}\\
			No.\ of slices & \multicolumn{2}{c}{$30-35$} & \multicolumn{2}{c}{39}\\
			\bottomrule
		\end{tabular}
	\label{stab:acqparam}
	
	\end{table}
\end{document}